\documentclass{article}
\usepackage{spconf,amsmath,graphicx}
\usepackage{multirow}
\usepackage[margin=1in]{geometry}
\usepackage{amssymb}
\usepackage{amsthm}
\usepackage{enumitem}
\usepackage{lipsum}

\newcommand{\newterm}[1]{{\bf #1}}
\DeclareMathOperator{\rank}{rank}

\setlist{nosep, leftmargin=14pt}

\usepackage{mwe} 

\newcommand\blfootnote[1]{%
  \begingroup
  \renewcommand\thefootnote{}\footnote{#1}%
  \addtocounter{footnote}{-1}%
  \endgroup
}

\title{Feature Gradient Flow for Interpreting Deep Neural Networks in Head and Neck Cancer Prediction}
 \name{Yinzhu Jin$^{\star}$ \qquad Jonathan C. Garneau$^{\dagger}$ \qquad P. Thomas Fletcher$^{\ddagger, \star}$}

 \address{$^{\star}$ Department of Computer Science, University of Virginia, Charlottesville, VA, USA\\
     $^{\dagger}$Department of Otolaryngology, University of Virginia, Charlottesville, VA, USA\\
     $^{\ddagger}$ Department of Electrical and Computer Engineering, University of Virginia,\\ Charlottesville, VA, USA\\}
\begin{document}
%
\maketitle
\begin{abstract}
This paper introduces feature gradient flow, a new technique for interpreting deep learning models in terms of features that are understandable to humans.
The gradient flow of a model locally defines nonlinear coordinates in the input data space representing the information the model is using to make its decisions.
Our idea is to measure the agreement of interpretable features with the gradient flow of a model.
To then evaluate the importance of a particular feature to the model, we compare that feature's gradient flow measure versus that of a baseline noise feature. We then develop a technique for training neural networks to be more interpretable by adding a regularization term to the loss function that encourages the model gradients to align with those of chosen interpretable features.
We test our method in a convolutional neural network prediction of distant metastasis of head and neck cancer from a computed tomography dataset from the Cancer Imaging Archive.
\end{abstract}
%
%

\blfootnote{

Copyright \copyright 2022 IEEE

Personal use of this material is permitted. Permission from
IEEE must be obtained for all other uses, in any current or future media, including reprinting/republishing this material for
advertising or promotional purposes, creating new collective
works, for resale or redistribution to servers or lists, or reuse
of any copyrighted component of this work in other works.

Published in: 2022 IEEE 19th International Symposium on Biomedical Imaging (ISBI)}

\section{Introduction}
Deep neural networks (DNNs) have great promise for predicting disease outcomes from medical imaging. For example, researchers have demonstrated that DNNs can predict outcomes in head and neck cancer, such as whether the cancer will metastasize, from computed tomography (CT) images with high accuracy~\cite{diamant2019deep,Kann2019}.
However, full adoption of such models is held back by the fact that DNNs are typically ``black boxes'', i.e., the image features that they learn and use to make predictions are not known or not interpretable by humans. This in turn leads to a lack of trust in DNNs by potential clinical users.

Several methods have been proposed for interpretable or explainable deep learning~\cite{gilpin2018explaining,aladi2018peeking} and applied in medical image analysis. The most commonly used methods fall into the class of saliency maps, e.g., Grad-CAM~\cite{selvaraju2017grad}. In these methods, information derived from the classifier gradient is displayed on an input image, highlighting where in the image the classifier is using information to make its decision. While these methods explain {\em where} a classifier is focusing, they do not explain {\em what} information it is using to make a decision. One other way is to build a decision tree that approximates the deep learning model~\cite{zhang2019interpreting}, which makes use of the naturally interpretable architecture of decision trees. Similar to saliency mapping methods, it highlights object parts that are important for decision making without showing the exact features the model relies on. Another class of methods, such as LIME~\cite{ribeiro2016should}, fit a linear approximation to a DNN in a local region of the input space. The idea is that the approximating linear classifiers can be interpreted more easily because they are given by a single parameter vector that provides a weighting of the importance for each input feature. Testing with concept activation vectors (TCAV)~\cite{kim2018interpretability} extended the linear approximation approach by fitting linear classifiers of binary concepts to the intermediate layer activations of a DNN. This direction was subsequently extended to linear regression of continuous concepts and applied in medical imaging tasks by Graziani et al.~\cite{Graziani2020}. A limitation of these concept attribution approaches is that they rely on a {\em linear} approximation to the activations of a DNN, whereas the flexibility of deep models comes from the fact that they are highly nonlinear.

In this paper, we seek to understand the nonlinear features that are being utilized in the decision process of a DNN. We do this by first developing the {\em fibered manifold geometry} that a classifier induces on the input space, decomposing it into nonlinear dimensions that are either relevant or irrelevant to the classifier's decision (Section \ref{sec:fiber}). Second, we develop a method for measuring the alignment of a given set of interpretable features along the gradient flow of this classifier geometry (Section \ref{sec:methods}). In this way we query to what extent the classifier is using information that is human interpretable. Furthermore, we develop a regularlization term for training a classifier to prefer using interpretable feature dimensions. Finally, we demonstrate the effectiveness of our approach in a prediction of distant metastasis from CT images of head and neck tumors (Section \ref{sec:results}). We show that our training method leads to a classifier that uses a higher percentage of interpretable image features.

\begin{figure}
    \centering
    \includegraphics[width=0.8\columnwidth]{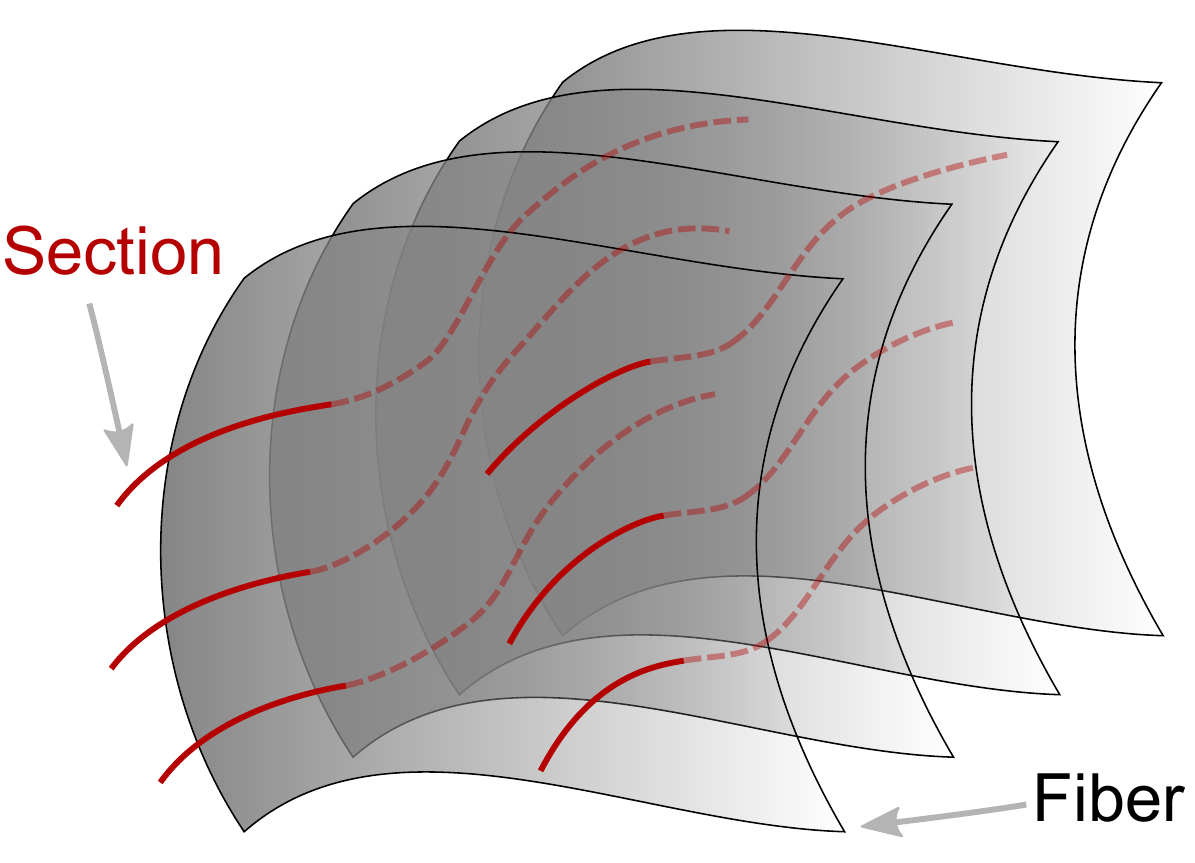}
    \caption{Fibered manifold geometry of a classifier.}
    \label{fig:fibers}
\end{figure}
\section{The Fibered Manifold Geometry of a Classifier}
\label{sec:fiber}
In this section we describe how a classifier locally decomposes the input data space into nonlinear dimensions along which the classifier decision changes, called \newterm{sections}, and complementary dimensions along which the classifier decision remains constant, i.e., the level sets of the classifier function, called \newterm{fibers}. This structure is known as a \newterm{fibered manifold} (see Fig.~\ref{fig:fibers}). The key insight to our approach is that to interpret a classifier one must understand the section dimensions because these are the dimensions along which the classifier is differentiating different classes of data. A classifier is ``ignoring'' the features along its fiber dimensions.

Let's consider a classifier taking inputs $x \in \mathbb{R}^d$ and predicting an
output class $y \in \{1, \ldots, K\}$. Furthermore, assume this classifier is a
$C^1$ mapping, $f : \mathbb{R}^d \rightarrow \mathbb{R}^K$. For example, the
outputs could be conditional probabilities, $p(y \mid x)$, or normalized logits,
$z = \ln(p(y \mid x))$. We will denote the Jacobian matrix of $f$ at a point
$x \in \mathbb{R}^d$ as $Df(x)$. The \newterm{rank} of $f$ at $x$ is defined as
$\rank(Df(x))$.

Assuming that $K < d$, the maximal rank of $f$ at any point is $K - 1$, due to the constraint that $\sum_y p(y \mid x) = 1$. A \newterm{regular point} of $f$ is a point $x \in \mathbb{R}^d$ such that $Df(x)$ has maximal rank, that is, $\rank(Df(x)) = K - 1$.
The set of regular points of $f$ is open in $\mathbb{R}^d$. This implies that there is a neighborhood about any regular point of $f$ that is a \newterm{fibered manifold}, i.e., there is a (possibly nonlinear) coordinate system that decomposes into $d - K + 1$ fiber coordinates, where $f$ remains constant, and $K-1$ section coordinates, where $f$ changes its output.




\section{Interpretable feature alignment}
\label{sec:methods}
In addition to our classifier function, $f : \mathbb{R}^d \rightarrow \mathbb{R}^K$, assume that we can also compute a set of $m$ {\em interpretable} features through a mapping $g : \mathbb{R}^d \rightarrow \mathbb{R}^m$. The general idea of our method is to measure how well the classifier $f$ is using these interpretable features by looking at the alignment of their section subspaces, e.g., if the dimensions in which they vary are similar.

Consider first the simple case where $K = 2$ (a binary classifier) and $m = 1$ (a single interpretable features). Then the sections defined by $f$ and $g$ are one dimensional and tangential to their respective gradients. Thus we can measure the agreement of the features by the alignment between the gradients of $f$ and $g$, e.g., the angle between them. We can estimate the expectation of this alignment over the data distribution by summing this value at each point in the test data. At a single data point $x \in \mathbb{R}^d$, this \newterm{pointwise alignment} is given by
\begin{equation}
\label{eq:atpoint}
    S(x) = \left(\frac{\langle \nabla f(x),\nabla g(x)\rangle}{\|\nabla f(x)\|\cdot\|\nabla g(x)\|}\right)^2.
\end{equation}
We square the dot product between the normalized gradients because we are only concerned about how well these dimensions align. We do not care about the magnitude of the units or the polarity of the gradients, that is, we consider the alignment of the bidirectional lines defined by the two gradients.


\subsection{Decomposition into Feature Hyperplane}
Now consider we want to evaluate the classifier's dependency on multiple features. We may derive this as a search for a classifier, $\phi : \mathbb{R}^m \rightarrow \mathbb{R}^K$ from the interpretable features, $g(x)$, that approximates the prediction output by $f$, i.e.,
\begin{equation}
\label{eq:phi}
    f(x) \approx \phi\circ g(x).
\end{equation}
If the task is difficult enough, then the same classification from $f$ will not be able to be computed using only the interpretable features $g(x)$. Therefore, the relationship in \eqref{eq:phi} is not an exact equality. This is what we expect in the case of a DNN, that is to say, we don't expect that the decision of a DNN can be explained perfectly by interpretable features alone. Rather than optimize for $\phi$ directly, we consider minimizing the difference in the gradients of both sides of~\eqref{eq:phi}. In other words, we want to minimize $\|\nabla f-Dg^T\nabla\phi\|$.
This gives us
\begin{equation*}
    \nabla \phi = (Dg Dg^T)^{-1} Dg \nabla f,
\end{equation*}
which can be viewed as a projection of $\nabla f$ to the hyperplane given by the rows of $Dg$, i.e., the gradients of the interpretable features. Then we can decompose $\nabla f$ into a parallel part and a vertical part:
\begin{align}
    \nabla f_{\|} &= Dg^T (Dg Dg^T)^{-1} Dg \nabla f,\label{eq:para}\\
    \nabla f_{\perp} &= \nabla f - \nabla f_{\|}.\label{eq:perp}
\end{align}
These can be understood as the interpretable component~\eqref{eq:para} and un-interpretable component~\eqref{eq:perp} of the classifier gradient using the selected features.

Note that
    $\|\nabla f\|^2 = \|\nabla f_{\|}\|^2 + \|\nabla f_{\perp}\|^2$,
so we can naturally define the alignment measure as the fraction of the squared gradient norm contained in tangent space to the section of the interpretable features, i.e.,
\begin{equation*}
    S(x) = \frac{\| \nabla f_{\|}(x)\|^2}{\| \nabla f(x) \|^2}.
\end{equation*}
Note that in the case with only one feature, this definition of $S$ corresponds to mission from IEEE must be obtained for all other uses, in any current or future media, including reprinting/republishing this material for advertising or promotional purposes, creating new collective works, for resale or redistribution to servers or lists, or reuse of any copyrighted component of this work in other works.”the previous one in \eqref{eq:atpoint}.

\subsection{Gradient Flow to the Decision Boundary}
Although observing the alignment at a point can give some indication whether the classifier is making use of the given features, it is only a local property and does not take into account more global geometry of the classifier. To address this, we next develop a measurement of how the gradients of interpretable features align with those of the classifier along a path from the data point to the classifier's decision boundary.

In practice, we can start from the data point and follow the gradient of the classifier and stop when it hits the decision boundary. Then we can integrate the normalized dot product of the gradients of the classifier and the gradients of the feature mapping along this path. To do this, first define the {\em gradient flow} from a data point $x$ as a curve in the data space $\gamma : [0, T] \rightarrow \mathbb{R}^d$ that begins at $\gamma(0) = x$ and follows the gradient of the classifier, i.e.,
\begin{equation}
\frac{d \gamma(t)}{dt} = \nabla f(\gamma(t)).    
\end{equation}
With a similar decomposition as the pointwise case, we measure the total fraction of alignment of the classifier and the interpretable features along the gradient flow asmission from IEEE must be obtained for all other uses, in any current or future media, including reprinting/republishing this material for advertising or promotional purposes, creating new collective works, for resale or redistribution to servers or lists, or reuse of any copyrighted component of this work in other works.”
\begin{equation*}
    F(x) = \frac{\int_0^T \left\| \nabla f_{\|}\left(\gamma(t)\right)\right\|^2 dt}{\int_0^T \left\| \nabla f\left(\gamma(t)\right) \right\|^2 dt}.
\end{equation*}
We note that both the pointwise score $S$ and the gradient flow score $F$ range from 0 to 1.

\subsection{Enhancing Interpretability During Training}
With the above measure of alignment, we can further use it to encourage a model to be more interpretable. This is done by adding an alignment ``reward'' term to the loss during training. Given interpretable features, $g_i(x), i = 1, \ldots, m,$ for each data point $x$, the training loss is:
\begin{equation}
\begin{aligned}
    \mathcal{L}(x, f, g) = &\sum_x  L(f(x),y)\\
    -&\sum_x \sum_{i=1}^m{\lambda_i\left( \frac{\left<\nabla f(x), \nabla g_i(x) \right>}{\|\nabla f(x)\|\cdot\|\nabla g_i(x)\|}\right)^2}
\end{aligned}
\end{equation}
where $L$ is the original training loss function, and $\lambda_i$ is a positive scalar parameter that controls the weight of the alignment with the $i$th interpretable feature. It rewards the gradients of the model for aligning with the gradients of given features. We can tune $\lambda_i$ to be as large as is possible without hurting the performance of the model.

\section{Results}
\label{sec:results}


\begin{figure}
    \centering
    \includegraphics[width=0.48\columnwidth]{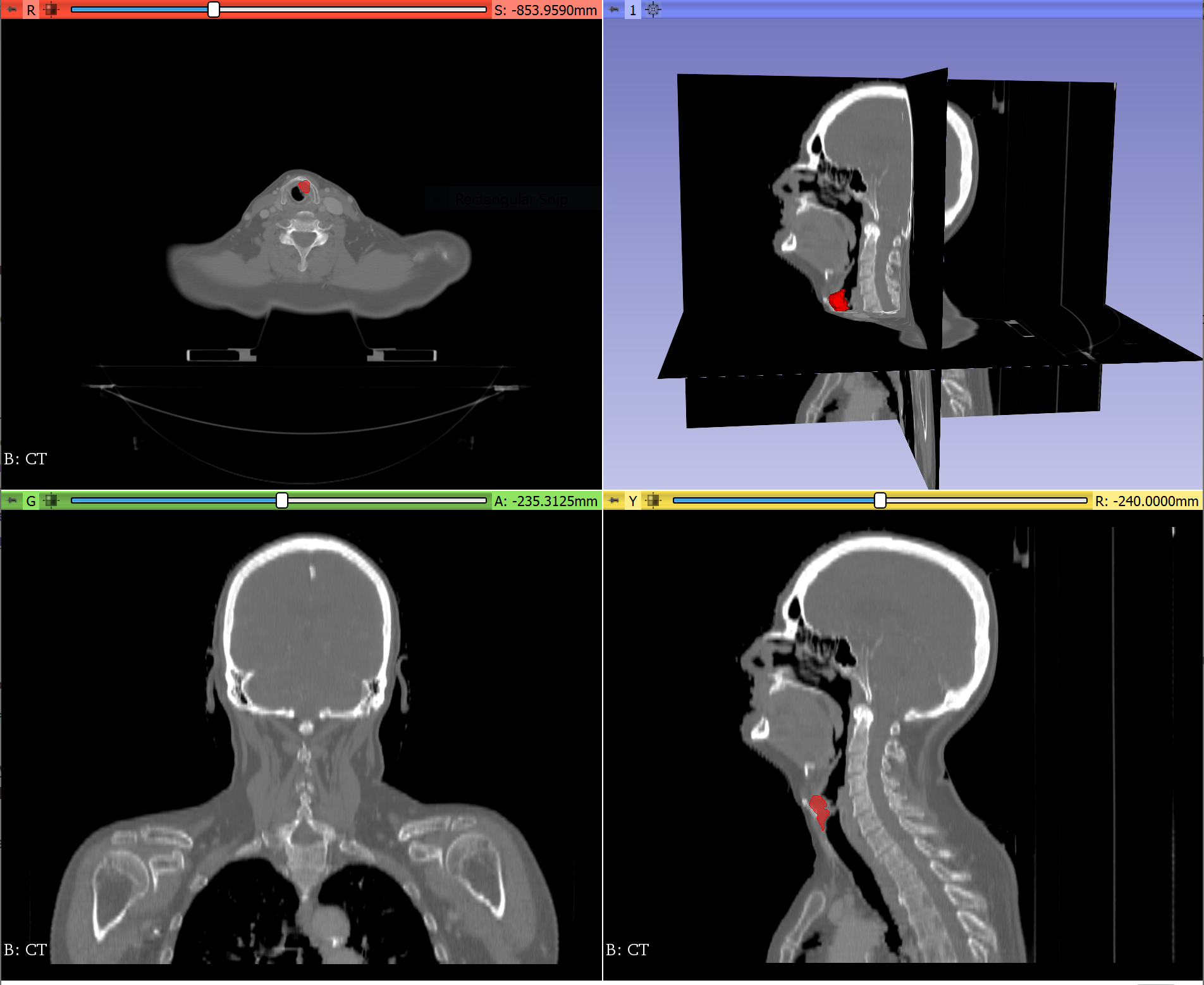}
    \includegraphics[width=0.48\columnwidth]{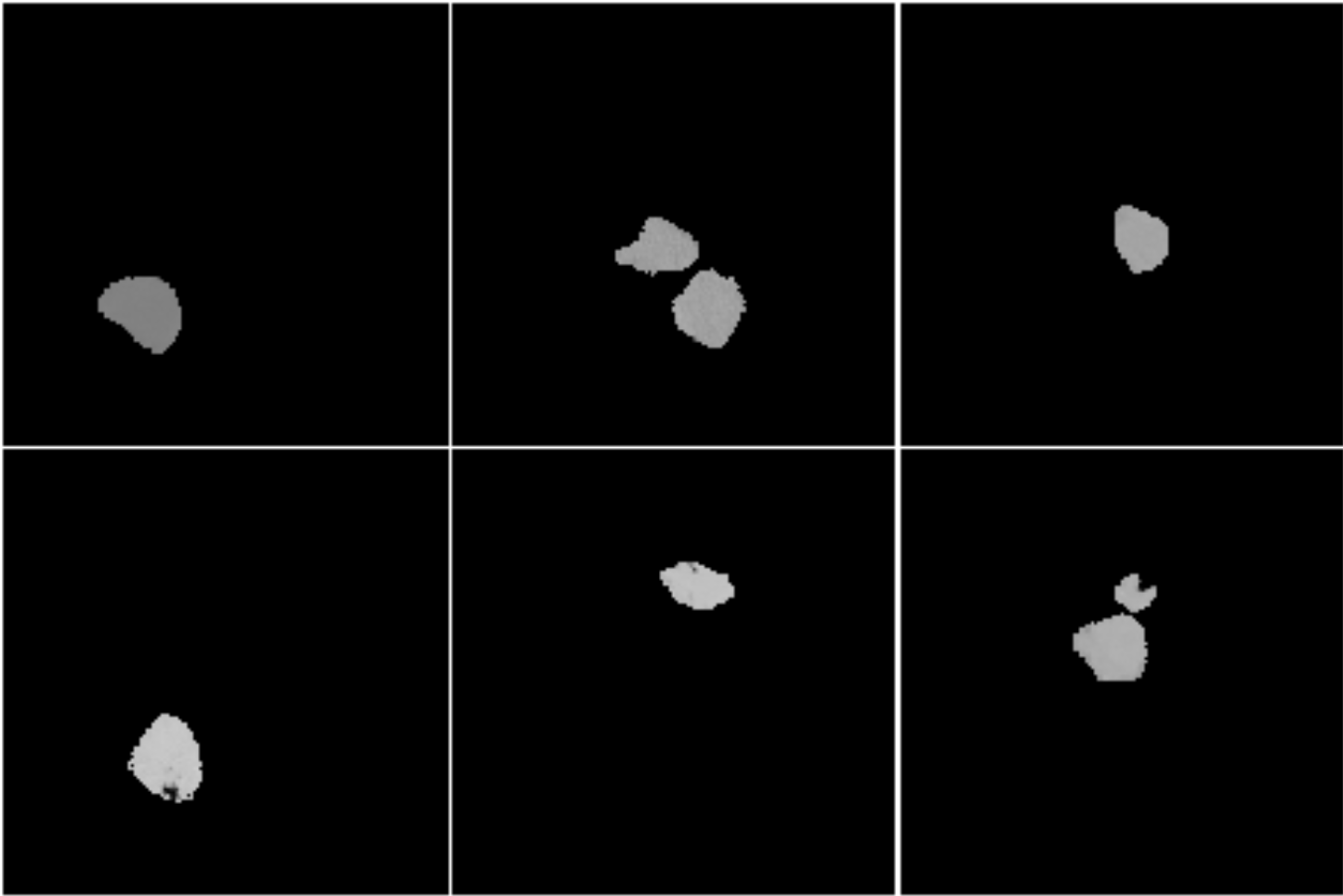}
    \caption{Example CT image with segmented tumor (left) and the masked tumor image used as input data (right).}
    \label{fig:data}
\end{figure}

\subsection{Dataset and Architecture}
We used a Head and Neck PET-CT dataset from \cite{vallieres2017radiomics} available on the Cancer Image Archive (TCIA) to evaluate our methods. The dataset includes data from 298 subjects. The task is to predict distant metastasis of head and neck tumors using CT images and segmentations of gross tumor volumes. It's a highly imbalanced classification task, with only 40 positive cases (13\%) in the entire dataset.

The classifier we used is a neural network with 3 convolutional layers (kernel sizes = $5 \times 5$, $3 \times 3$, and $3 \times 3$) each followed by average pooling and exponential linear units (ELUs). These are followed by 3 fully-connected layers (output dimensions = 256, 128, and 1). We chose average pooling over max pooling and ELU over ReLU because they are differentiable, while providing equivalent classification performance. The input data is $512\times 512$ gray scale images.

Following the work from Diamant et al.~\cite{diamant2019deep}, the inputs are 2D CT slices, each chosen where the cross-sectional area of tumor is the largest. The area outside the tumor was masked as zero. We augmented the data by randomly rotating in the range $\pm 20$ degrees and translating in the range $\pm 0.015$ times the image width/height. We used weighted random sampling to balance the training batches evenly between negative and positive samples.
The data was split into training set of 209 samples and test set of 89 samples. The model was trained 100 epochs using Adam optimizer with initial learning rate $3.0 \times 10^{-5}$ and batch size 32. For the model with alignment term, we chose 3 features each with with $\lambda_i = 3.0 \times 10^{-5}$. Our plain classifier achieved 0.681 balanced accuracy, and the classifier with alignment term training achieved 0.688 balanced accuracy.

\subsection{Interpretable Features}
We chose 3 features that can be calculated from the data, namely, overall brightness ($g_1$), tumor extent ($g_2$), and log aspect ratio of tumor ($g_3$). Let $I(u)$ denote an image, with pixel grid coordinates $u = (u_1, u_2)$. Then the three features are calculated as follows:
\begin{equation*}
\begin{aligned}
    g_1 &= \sum_u I(u)\\
    \mu &= \frac{1}{g_1} \sum_u u I(u),\,\, C = \frac{1}{g_1} \sum_u I(u) (u - \mu)(u - \mu)^T\\
    g_2 &= \mathrm{tr}(C)\\
    g_3 &= \log(\sigma_1) - \log(\sigma_2)\\
\end{aligned}
\end{equation*}
where $\sigma_1^2 \geq \sigma_2^2$ are eigenvalues of the covariance, $C$.

\begin{table}[h]
    \centering
    \caption{Results of Pointwise Alignment ($S$)}
    \label{tab:pointwise}
    \begin{tabular}{|c|c|c|c|}
    \hline
        \multirow{2}{4em}{feature} & \multicolumn{2}{|c|}{Means} & \multirow{2}{3em}{$p$-value}\\
        \cline{2-3}
        & plain & enhanced & \\
        \hline
         overall brightness & 8.2e-4 & 3.4e-3 & $<10^{-6}$ \\
         tumor extent & 3.8e-2 & 2.4e-2 & ---\\
         log aspect ratio & 1.3e-2 & 9.5e-3 & --- \\
         combined features & 0.12 & 0.16 & $<10^{-6}$ \\
         \hline
         single random & 3.6e-6 & 4.1e-6 & 0.95\\
         three random  & 1.2e-5 & 1.3e-5 & 0.30\\
         \hline
    \end{tabular}
\end{table}

\subsection{Interpretability Measures}
Here we apply our pointwise interpretability measure, $S$, to our test set. The results are shown in Table~\ref{tab:pointwise} for the both the plain model and the model trained with our enhanced interpretability method. While the interpretability scores for the individual features seem quite small, we show that they are actually large relative to the interpretability scores for a randomly generated feature (which provides a baseline interpretability score for a feature that should {\em not} be useful to the classifier). We generated one random feature from a standard normal distribution to compare to the single feature case, and three independent random features to compare to the multiple feature case. These two are referred as ``single random'' and ``three random'' in the table. To quantify if our interpretable features were statistically significantly better than random to the classifier, we performed a Kolmogorov-Smirnov (KS) test between the distribution of $S$ values. All three features $g_1, g_2, g_3$, and their combination, were statistically significantly better than random at $p < 10^{-3}$. The last column in Table~\ref{tab:pointwise} is the KS test $p$-value to see if the enhanced training improved the interpretability over the plain model. As we can see, tumor extent and aspect ratio did not become more useful to the classifier, but brightness did. Finally, using all three interpretable features jointly accounts for 16\% of the classifier's squared gradient magnitude.
From the results for the model with alignment term, we can see that the interpretable fraction of classifier gradients increased while not negatively affecting classifier performance.
\begin{table}[h]
    \centering
    \caption{Results of Gradient Flow Alignment ($F$)}
    \label{tab:flow}
    \begin{tabular}{|c|c|c|c|}
    \hline
        \multirow{2}{4em}{feature} & \multicolumn{2}{|c|}{Means} & \multirow{2}{3em}{$p$-value}\\
        \cline{2-3}
        & plain & enhanced & \\
        \hline
         overall brightness & 8.4e-4 & 3.4e-3 & $<10^{-6}$\\
         tumor extent & 3.4e-2 & 2.1e-2 & ---\\
         log aspect ratio& 4.1e-3 & 2.0e-3 & --- \\
         combined features & 0.14 & 0.19 & $<10^{-6}$ \\
         \hline
         single random & 4.4e-6 & 3.8e-6 & 0.40\\
         three random & 1.1e-5 & 1.2e-5 & 0.95 \\
         \hline
    \end{tabular}
\end{table}

We also show in Table~\ref{tab:flow} the results of the gradient flow alignment measure, $F$, applied on the both the plain model and the model trained with our interpretability enhancement. The overall behavior is similar to that of the pointwise interpretability scores. Again, the KS test indicates that the gradient flow interpretability measures for all three features $g_1, g_2, g_3$, and their combination, were statistically significantly better than random at $p < 10^{-3}$. Interestingly, the gradient flow measures are similar to the pointwise measures for individual features, but somewhat higher for the three features combined.

\section{Discussion}
We introduced a new method to evaluate the importance of given interpretable features by quantifying their gradient alignment along the gradient flow of the model. 
Although the resulting alignment scores may seem small, we note that they are significantly larger than random and account for a high proportion of the variance relative to the high dimensionality of the input images. 
A limitation of our method is that it requires the user to input interpretable features of interest. Thus, our method may be less effective in cases where a model mostly makes use of unexpected features.

\section{Compliance with Ethical Standards}
This research study was conducted retrospectively using human subject data made available in open access by Vallieres et al.~\cite{vallieres2017radiomics} in the Cancer Image Archive (TCIA). Ethical approval was not required as confirmed by the license attached with the open access data.

\bibliographystyle{IEEEbib}
\bibliography{refs}

\end{document}